\documentclass[twocolumn,pre,aps,10pt,showpacs,superscriptaddress]{revtex4-1}

\usepackage{graphicx}
\usepackage{amssymb,amsfonts,amsmath}

    \usepackage{multirow}

\usepackage{slashbox}

\begin{document}

\title{The crystallization of asymmetric patchy models for globular proteins in solution}

\author{Diana Fusco}
\affiliation{Program in Computational Biology and Bioinformatics, Duke University, Durham, NC 27708}

\author{Patrick Charbonneau}
\affiliation{Program in Computational Biology and Bioinformatics, Duke University, Durham, NC 27708}
\affiliation{Department of Chemistry, Duke University, Durham, NC 27708}
\affiliation{Department of Physics, Duke University, Durham, NC 27708}

\begin{abstract} 
Asymmetric patchy particle models have recently been shown to describe the crystallization of small globular proteins with near quantitative accuracy. Here, we investigate how asymmetry in patch geometry and bond energy generally impact the phase diagram and nucleation dynamics of this family of soft matter models. We find the role of the geometry asymmetry to be weak, but the energy asymmetry to markedly interfere with the crystallization thermodynamics and kinetics. These results provide a rationale for the success and occasional failure of George and Wilson's proposal for protein crystallization conditions as well as physical guidance for developing more effective protein crystallization strategies.
\end{abstract}

\maketitle

\section{Introduction}

Proteins are key biological molecules whose physiological roles are, for the most part, tightly linked to their three-dimensional structure. 
Because X-ray and neutron crystallography are the most widely used techniques to detail these structures~\cite{mcpherson:1999,Chayen2009}, the difficulty of obtaining diffraction-quality protein crystals severely limits our understanding of living systems. 
Crystallizing a protein typically involves placing a drop of protein solution near a high-salt aqueous buffer that drives the vapor diffusion away from the drop. The non-volatile solutes then steadily concentrate and, if the initial conditions are properly chosen, a protein crystal assembles~\cite{mcpherson:1999}.  From a physical point of view, identifying successful crystallization conditions is thus equivalent to determining the protein's solution phase diagram. The limited usefulness of existing physical descriptions and of knowledge-based approaches~\cite{george:1994,derewenda:2004}, however, leaves a vast space of experimental conditions to be screened. A material understanding of protein assembly is thus essential to developing more effective crystallization strategies.

Soft matter descriptions of protein assembly based on particles with isotropic, short-range attractive interactions~\cite{Hagen1994,rosenbaum:1996,Wolde1997} -- as suggested by early structural studies~\cite{janin:1995, carugo:1997, bahadur:2004} --  provide some conceptual guidance. They identify the region between the solubility line, above which the solution is stable, and the liquid-liquid critical point, well below which the system precipitates into amorphous materials~\cite{Lu2008,Fortini2008}, as the ``crystallization gap'' where crystal assembly is possible. This schematic picture is, however, unable to reproduce many experimental trends~\cite{haas:1999,lomakin:1996,Lomakin1999,curtis:2001,mcmanus:2007}. The introduction of bond directionality in symmetric ``patchy'' models aimed to better represent the effective protein-protein interactions that drive their crystallization~\cite{Gogelein2008, Bianchi2011}. Yet the most commonly studied versions of these models have symmetric and interchangeable patches, which are atypical of real proteins~\cite{X-ray,derewenda:2004,Pellicane2008,Dorsaz:2012} and insufficient to describe the assembly of even the simplest of globular proteins~\cite{Dorsaz:2012,Fusco2012}. 
In this article, we investigate the role of patch geometry and bond energy asymmetry on the phase diagram and assembly dynamics of a coarse-grained protein model of rigid globular proteins in aqueous solution. 
This additional anisotropy `direction' complements earlier experimental proposals~\cite{glotzer:2007}. It also maps onto the assembly of more complex structures in systems such as DNA-coated colloidal particles, in which the strength of directional interaction can be finely tuned~\cite{glotzer:2004,frenkel:2011,Pine:2012}.   Our work therefore identifies general regions of parameter space that should be targeted for specific colloidal assemblies, such as gel and crystal formation.

The plan for the paper is as follows. In Section II, we describe the model used. In Section III, we analyze the phase diagrams of a collection of model parameters. In Section IV, we study nucleation and the pathways to crystallization. Finally in Section V, we determine how percolation interferes with crystallization.

\section{Model description} 

We describe each protein as a hard sphere of diameter $\sigma$, which sets the unit of length, with interacting directional patches that mimic the effective interactions between solvated proteins. This schematic description assumes that proteins maintain their structure throughout crystallization and that solvated electrolytes screen long-range electrostatic interactions, which is typical of protein solutions that produce diffraction quality crystals~\cite{dale:2003,slabinski:2007}. Crystallization cocktails that include salt as only cosolute account for nearly 50\% of successful experimental conditions in typical databases~\cite{Charles2006}. In these conditions, attraction is triggered by the specific chemical details at the protein surface and thus directional interactions dominate. This treatment complements and supports with previous studies that focused on the interplay between specific and non-specific (depletion) interactions~\cite{Whitelam2010,Haxton2012,Haxton2013}. 

We consider a variant of the patchy model of Ref.~\cite{Russo2011} in which patch-patch interactions are specific~\cite{Sear1999} and their range and strength are independently tunable~\cite{Kern2003}.
The pair-wise interaction between particles 1 and 2, whose centers are a distance $r_{12}$ apart, is
\begin{eqnarray}
\phi(r_{12},\Omega_1,\Omega_2)&=&\phi_{\mathrm{HS}}(r_{12})+\sum _{i=1}^n[\phi_{2i,2i-1}(r_{12},\Omega_1,\Omega_2)\nonumber\\
&&+\phi_{2i-1,2i}(r_{12},\Omega_1,\Omega_2)],
\end{eqnarray}
where $\Omega_1$ and $\Omega_2$ are the Euler angles and $n$ is the number of pairs of patches. A hard-sphere (HS) potential captures volume exclusion
\begin{equation}
\phi_{\mathrm{HS}}(r)=\left\{
\begin{array}{cc}
\infty&r\leq\sigma\\
0&r>\sigma.
\end{array}\right .
\end{equation}
The patch-patch interaction is the product of radial and an angular components
 \begin{equation}
 \phi_{2i,2i-1}(r_{12},\Omega_1,\Omega_2)=\psi_{i}(r_{12})\omega_{2i,2i-1}(\Omega_1,\Omega_2),
 \end{equation}
 where
 \begin{equation}
\psi_{i}(r)=\left\{
\begin{array}{cc}
-\epsilon_i&r\leq\lambda_i\sigma\\
0&r>\lambda_i\sigma
\end{array}\right .,
\end{equation}
and
\begin{equation}
\omega_{2i,2i-1}(\Omega_1,\Omega_2)=\left\{
\begin{array}{cc}
1&\theta_{1,2i}\leq\delta_{2i} \hbox{ and } \theta_{2,2i-1}\leq\delta_{2i-1}\\
0&\hbox{otherwise}
\end{array}\right . .
\end{equation}
The interaction range $\lambda_i$ is in units of $\sigma$, $\delta_{2i}$ is the semi-width of  patch $2i$ and $\theta_{1,2i}$ is the angle between the vector $\mathbf{r}_{\mathbf{12}}$  and the vector defining patch $2i$ on particle $1$. By symmetry an analogous definition holds for $\theta_{2,2i-1}$. 
 Here, the short radial extent of the square-well attraction, $\lambda_i=1.1\sigma$~\cite{noro:2000}, and its surface coverage measured by the semi-opening angle of its conical segment, $\delta_i=\cos^{-1}(0.89)$, are chosen to be typical of protein-protein interactions~\cite{Pellicane2008,Fusco2012}. 
By contrast, the patch position on the surface and the bond energy $\epsilon_i$ are randomly chosen, under the sole constraint that the lattice formed by simply bonding the patches is that of the most commonly observed in monomeric protein crystals, the orthorhombic $P2_12_12_1$~\cite{footnote:1,Wukovitz1995}. This lattice's three non-intersecting two-fold screw axes guarantee a high number of rigid-body degrees of freedom with minimal symmetry constraints. 

\begin{figure}[tbh]
\begin{center}
\includegraphics[width=0.4\textwidth]{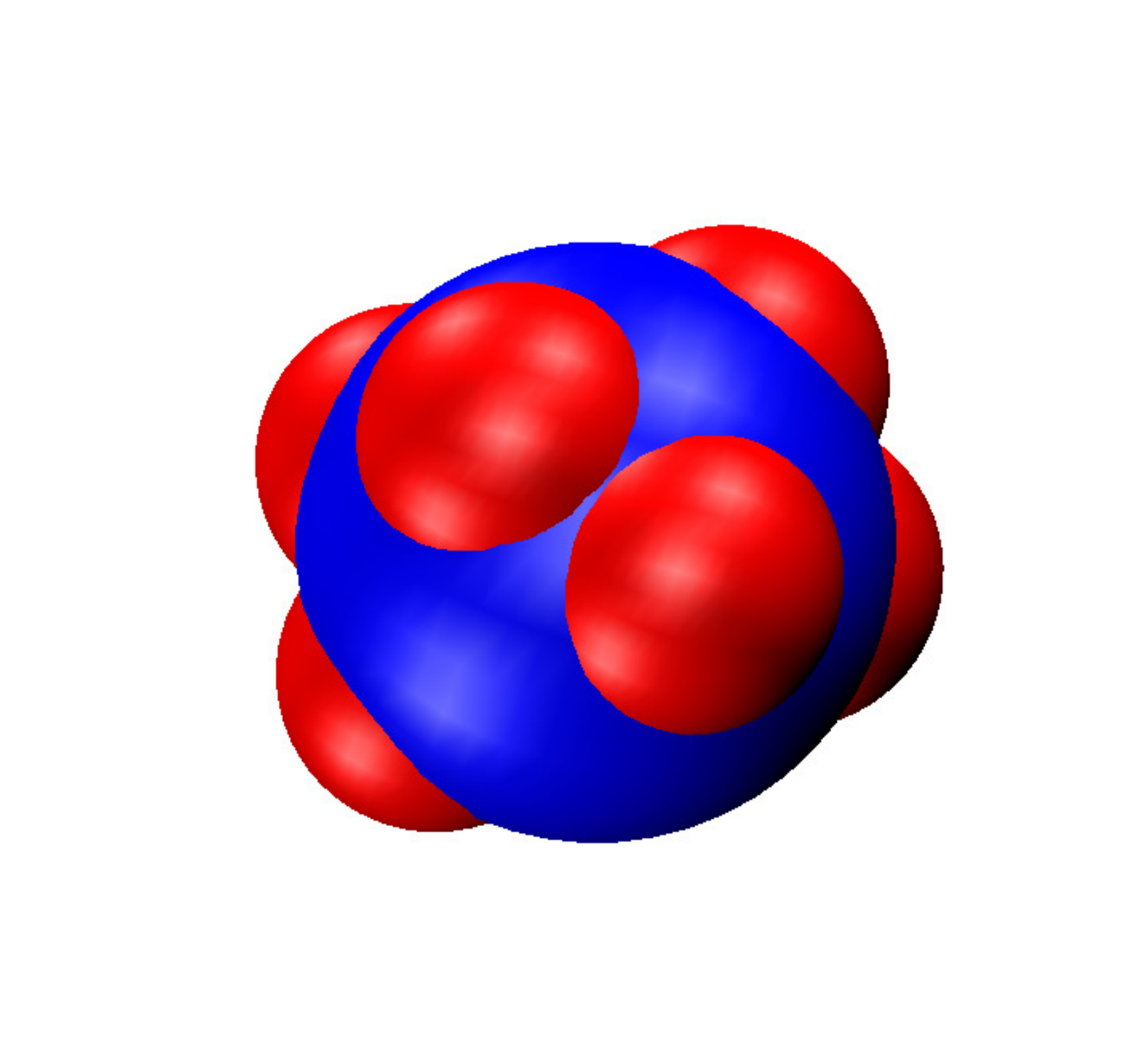}
\caption{(Color online) Sketch of a patchy particle. To determine the inertia tensor, we treat the patches as spherical balls (red/light gray) whose center is at the particle surface.}\label{fig:inertia}
\end{center}
\end{figure}

We summarize the patch properties with energy and geometry asymmetry parameters
\begin{align}
\zeta&=\frac{(\epsilon_1-\epsilon_2)^2+(\epsilon_1-\epsilon_3)^2+(\epsilon_2-\epsilon_3)^2}{2(\epsilon_1^2+\epsilon_2^2+\epsilon_3^2)}\nonumber\\
\gamma&=\frac{(I_1-I_2)^2+(I_1-I_3)^2+(I_2-I_3)^2}{2(I_1^2+I_2^2+I_3^2)}\nonumber,
\end{align}
where $I_i$ represents the $i^{th}$ eigenvalue of the inertia tensor of the object represented in Fig.~\ref{fig:inertia}. Each patch (in red) carries a mass $M$ at its center. The inertia tensor is computed over the set of weighted patches. The expression for $\gamma$ guarantees that its value does not depend on the fictitious mass $M$ nor on the radius of the particle, as long as they do not vary from one patch to the other~\cite{Miller2010}. Note that patches located on a perfect octahedron have $I_1=I_2=I_3$, and consequently $\gamma=0$.
Both $\zeta$ and $\gamma\in[0,1]$, where 0 corresponds to an equal energy distribution ($\epsilon_1$=$\epsilon_2$=$\epsilon_3$) and cubically distributed patches, and 1 corresponds to a complete energy asymmetry ($\epsilon_1$=$\epsilon_2$=0 and $\epsilon_3$=$\epsilon_{\mathrm{tot}}$) and a unit cell elongated in a single direction.
Because of the $P2_12_12_1$ constraint on the crystal, patches cannot be too close to one another. Otherwise, bonded particles would overlap and the unit cell would stretch beyond the range of attraction $\lambda=1.1\sigma$, which limits the achievable asymmetry and sets $\gamma\lesssim0.1$. Because a cubic symmetry ($\gamma=0$ and limiting case of $P22_12_1$) is not realizable within the three screw axes symmetry of $P2_12_12_1$, $\gamma$ is limited from below as well. The adopted range and width of the interaction and the $P2_12_12_1$ symmetry ensures that two particles can only interact one bond at a time. This condition, together with the impossibility for a patch to interact with its copy, also prevents dimerization. 
 Note that because the definition of $\gamma$ and $\zeta$ is purely geometrical, there is no reason to expect that different models with identical asymmetry parameters should behave identically. Tables~\ref{table:geo} and~\ref{table:energy} summarize the parameter values used in this work.
 
 \begin{table}[htb]
\centering
\caption{Geometry parameters: the triplets of numbers represent the unit vector coordinates of each patch. The center of patch 0 interacts with patch 1, patch 2 with patch 3 and patch 4 with patch 5. The first example $\gamma=0$ is reported for clarity.}\label{geometry}
\begin{tabular}{c|r@{.}lr@{.}lr@{.}lr@{.}lr@{.}lr@{.}l}
\hline
\multicolumn{1}{c|}{$\gamma$}
&\multicolumn{2}{c}{patch$_0$}
&\multicolumn{2}{c}{patch$_1$}
&\multicolumn{2}{c}{patch$_2$}
&\multicolumn{2}{c}{patch$_3$}
&\multicolumn{2}{c}{patch$_4$}
&\multicolumn{2}{c}{patch$_5$}\\
\hline
\multirow{3}{*}{0}&1&0&-1&0&0&0&0&0&0&0&0&0\\
&0&0&0&0&1&0&-1&0&0&0&0&0\\
&0&0&0&0&0&0&0&0&1&0&-1&0\\
\hline
\multirow{3}{*}{0.0172}&-0&8036&0&8036&-0&5186&-0&5186&0&3081&0&3081\\
&-0&5042&-0&5042&0&2731&0&2371&-0&8084&0&8084\\
&-0&3163&-0&3163&0&8102&-0&8102&0&5016&0&5016\\
\hline
\multirow{3}{*}{0.0217}&-0&7904&0&7904&-0&4227&-0&4227&0&3571&0&3571\\
&-0&5184&-0&5184&0&2807&0&2897&-0&776&0&776\\
&-0&3263&-0&3263&0&8617&-0&8617&0&5199&0&5911\\
\hline
\multirow{3}{*}{0.0381}&-0&7191&0&7191&-0&3813&-0&3813&0&3475&0&3475\\
&0&5659&0&5659&-0&1669&-0&1669&-0&7668&0&7668\\
&-0&4032&-0&4032&-0&9093&0&9093&0&5397&0&5397\\
\hline
\multirow{3}{*}{0.0631}&-0&6167&0&6167&0&6103&0&6103&-0&006&-0&006\\
&0&6521&0&6521&-0&3099&-0&3099&-0&9579&0&9579\\
&-0&441&-0&441&0&729&-0&729&0&2871&0&2871\\
\hline
\multirow{3}{*}{0.0787}&-0&9042&0&9042&-0&6276&-0&6276&0&2859&0&2859\\
&-0&3335&-0&3335&0&5386&0&5386&-0&9074&0&9074\\
&-0&2669&-0&2669&0&5622&-0&5662&0&3081&0&3081\\
\hline
\end{tabular}\label{table:geo}
\end{table}

\begin{table}[htb]
\centering
\caption{Energy parameters with $\epsilon_{\mathrm{tot}}=6$.}\label{energy}
\begin{tabular}{r@{.}l|r@{.}lr@{.}lr@{.}l}
\hline
\multicolumn{2}{c|}{$\zeta$}
&\multicolumn{2}{c}{$\epsilon_1$}
&\multicolumn{2}{c}{$\epsilon_2$}
&\multicolumn{2}{c}{$\epsilon_3$}\\
\hline
0&00&2&&2&&2&\\
0&11&1&2462&2&5482&2&2056\\
0&21&2&1&2&9066&0&9934\\
0&33&0&4854&2&8266&2&688\\
0&49&3&5756&0&2037&2&2207\\
0&50&3&&3&&0&\\
0&55&3&96&1&8&0&24\\
0&64&4&32&1&5&0&18\\
0&79&4&8&0&6&0&6\\
\hline
\end{tabular}\label{table:energy}
\end{table}

\begin{figure*}[tb]
\begin{center}
\includegraphics[width=0.9\textwidth]{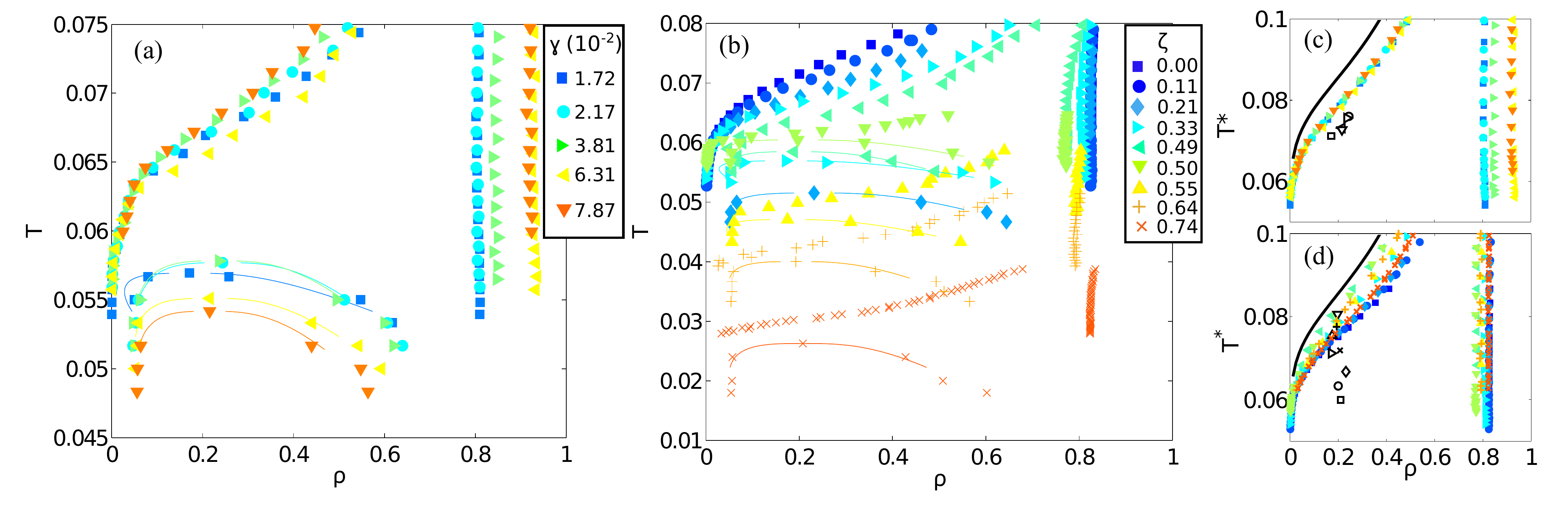}
\caption{(Color online) Temperature-density $\rho$ phase diagrams of patchy models with (a) different $\gamma$ at fixed $\zeta=0.33$, and (b) with different $\zeta$ at fixed $\gamma=0.0172$. (c) and (d) depict the same phase diagrams with $T$ rescaled following WPT. The crystal-fluid coexistence lines (symbols) are then close to, yet distinct from the WPT solubility line (solid black line). For visual clarity, only the gas-liquid critical points (black symbols) are reported in (c) and (d). Spontaneous crystallization during the simulations prevents the precise evaluation of the gas-liquid line for low $\zeta$. The specific parameter values are given in Tables~\ref{table:geo} and~\ref{table:energy}.}\label{fig:PD}
\end{center}
\end{figure*}

\section{Phase diagram} For these 30 randomly selected sets of patch geometry and bond energies we numerically determine the solubility line using free-energy integration and the metastable vapor-liquid line using Gibbs ensemble Monte Carlo simulations and compared the results to Wertheim's theory predictions.

\subsection{Phase diagrams from simulations}

Gibbs ensemble MC simulations (GEMC) directly determine the coexistence densities of the metastable gas and liquid phases \cite{Gibbs}. We simulate a total of $N=$1000 particles for $10^6$ MC cycles, where each cycle consists on average of $N$ particle displacements, $N$ particle rotations, $N/10$ particle swaps, and 2 volume $V$ changes. The critical temperature $T_c$ and density are then estimated using the law of rectilinear diameters \cite{frenkel:2001}. 

Because the gas-liquid line is metastable, for low energy asymmetry crystallization happens so quickly that determining the gas and liquid coexistence densities is impossible. In such cases, we estimate the critical temperature from Wertheim's perturbation theory (see below). 

To determine the fluid-solid coexistence curve,  we integrate the Clausius-Clapeyron equation starting from one coexistence point, using a fourth-order predictor-corrector algorithm~\cite{Hagen1994}.
The coexistence point itself is determined using free energy calculations. The free energy of the fluid is computed using thermodynamic integration from the free energy of an ideal gas~\cite{Vega2008}. The free energy of the crystal is determined using an Einstein crystal with a fixed center of mass as reference~\cite{frenkel1984}. Its Hamiltonian is
\begin{equation}
H^{\textrm{Ein}}(\Xi_{\textrm{trans}},\Xi_{\textrm{or}})=\Xi_{\textrm{trans}}\sum_{i=1}^N(\mathbf{r}_i-\mathbf{r}_{i,0})^2+\Xi_{\textrm{or}}\sum_{i=1}^Nf(\theta_i,\phi_i,\chi_i),\nonumber
\end{equation}
where $f(\theta_i,\phi_i,\chi_i)=1-\cos(\psi_{i,1})+1-\cos(\psi_{i,2})$ , $(\theta_i, \phi_i,\chi_i)$ are the Euler angles describing the orientation of particle $i$, and $\psi_{i,j}$ is the angle formed between the vector defining patch $j$ of particle $i$ and the corresponding vector in the Einstein crystal. As explained in Ref.~\cite{Vega2008}, the Helmhotz free energy of the reference Einstein crystal can then be written as
\begin{equation}
a^{\textrm{COM}}_{\textrm{Ein}}=a^{\textrm{COM}}_{\textrm{trans}}+a^{\textrm{COM}}_{\textrm{or}},
\end{equation}
where
\begin{equation}
\beta a^{\textrm{COM}}_{\textrm{trans}}=-\frac{3}{2}\frac{N-1}{N}\ln\left(\frac{\pi}{\beta\Xi_{\textrm{trans}}}\right)-\frac{3}{2N}\ln N
\end{equation}
and
\begin{equation}
\beta a^{\textrm{COM}}_{\textrm{or}}=-\ln\left\{\frac{1}{8\pi^2}\int{d\theta\sin(\theta)d\phi d\chi \exp[-\beta\Xi_{\textrm{or}}f(\theta,\phi,\chi)]}\right\}.\nonumber
\end{equation}
The calculation of $a^{\textrm{COM}}_{\textrm{trans}}$ is straightforward, but that of $a^{\textrm{COM}}_{\textrm{or}}$ requires either a tedious numerical integration or an analytical approximation. We opt for the latter using a saddle point approximation, which is accurate and efficient for the high values of $\beta\Xi_{\textrm{or}}$ used here, because the integrand is then sharply peaked. Defining $(\theta_0,\phi_0,\chi_0)$ as the reference orientation in the Einstein crystal and changing variable ${\boldsymbol \alpha}=(\cos(\theta),\phi,\chi)$  gives
\begin{align}
\int{d\theta\sin(\theta)d\phi d\chi \exp[-\beta\Xi_{\textrm{or}}f(\boldsymbol \alpha)]}&=\int d\boldsymbol \alpha \exp[-\beta\Xi_{\textrm{or}}f(\boldsymbol \alpha)]\nonumber\\
\approx\frac{\exp[-\beta\Xi_{\textrm{or}}f(\boldsymbol \alpha_0)](2\pi)^{3/2}}{(\beta\Xi_{\textrm{or}})^{3/2}\det(H[f(\boldsymbol \alpha_0)])^{1/2}}&\nonumber\\
=\frac{(2\pi)^{3/2}}{(\beta\Xi_{\textrm{or}})^{3/2}\det(H[f(\boldsymbol \alpha_0)])^{1/2}},\nonumber
\end{align}
such that
\begin{equation}\label{eq:saddle_point}
\beta a^{\textrm{COM}}_{\textrm{or}}\approx\frac{3}{2}\ln(\beta\Xi_{\textrm{or}})+\frac{1}{2}\ln\{8\pi \det(H[f(\boldsymbol \alpha_0)])\},
\end{equation}
where $\det(H[f(\boldsymbol \alpha_0)])$ is the determinant of the Hessian of function $f$ computed at $\boldsymbol \alpha_0$. Its analytical expression is reported in Appendix A. Once the free energy of the reference crystal is known, the free energy of the actual crystal is obtained following a standard free energy integration protocol~\cite{Romano2010}. Several simulations along an isobar starting from the fluid and from the crystal are then necessary to determine the temperature at which the chemical potential of the two phases coincides~\cite{Vega2008,Romano2010}.

Figure~\ref{fig:PD} illustrates the simulated phase diagrams. The gas-liquid critical temperature $T_c$ generally decreases with increasing $\gamma$ because patch proximity anti-correlates bond formation and decreases the liquid entropy, although the limited number of systems studied partially hides this feature (Table~\ref{table:Tc}). The solubility line, by contrast, is clearly similar for different geometries at fixed $\zeta$ and monotonically shifts to lower temperatures with increasing energy asymmetry. 

\begin{table}[htp]
\centering
\caption{Critical temperatures $T_c$ for the models studied. $^*$~indicates that the system crystallized spontaneously in GEMC simulations and the Wertheim's estimate is instead reported. -~indicates models for which the phase diagram was not determined. Temperatures are in units of $\epsilon_{\mathrm{tot}}$.}
\begin{tabular}{|c|ccccc|}
\hline
\backslashbox{$\zeta$}{$\gamma$}&0.0172&0.0217&0.0381&0.0631&0.0787\\
\hline
0.00$^*$&0.052&0.052&0.052&0.052&0.052\\
0.11$^*$&0.053&0.053&0.053&0.053&0.053\\
0.21&0.052&-&-&-&-\\
0.33&0.057&0.058&0.058&0.055&0.054\\
0.49&0.059&0.057&0.053&0.050&0.055\\
0.50&0.061&-&-&-&-\\
0.55&0.047&-&-&-&-\\
0.64&0.040&-&-&-&-\\
0.79&0.026&0.024&0.024&0.019&0.020\\
\hline
\end{tabular}\label{table:Tc}
\end{table}

\subsection{Phase diagrams from Wertheim's perturbation theory}

According to Wetheim's perturbation theory~\cite{wertheim:1984a,wertheim:1984b}, the fluid free energy can be approximated by the hard sphere free energy plus a bond free energy correction 
\begin{equation}
a_f=a_{\mathrm{HS}}+a_{\mathrm{bond}},
\end{equation}
where
\begin{equation}
\beta a_{\mathrm{bond}}=\sum_{a\in\Gamma}\left(\ln X_a-\frac{X_a}{2}\right)+\frac{m}{2}.
\end{equation}
Here $m$ is the total number of attractive sites, $X_a$ is the probability that the molecule is not bonded at site $a$, and $\Gamma$ is the set of interacting patches. 

Similarly, the chemical potential is given by
\begin{equation}
\beta\mu_f=\beta a_f+\frac{\beta p}{\rho}=\beta a_{\mathrm{HS}}+\beta a_{\mathrm{bond}}+\frac{\beta p_{\mathrm{HS}}}{\rho}+\frac{\beta p_{\mathrm{bond}}}{\rho},
\end{equation}
where the pressure $p$ contribution to bonding is
\begin{equation}
\beta p_{\mathrm{bond}}=\rho^2\sum_{a\in\Gamma}\left(\frac{\partial X_a}{\partial\rho}\right)\left(\frac{1}{X_a}-\frac{1}{2}\right).
\end{equation}

In the solid, $\beta a_s\approx\beta\mu_s$, because the ratio $\frac{\beta p_s}{\rho_s}$ is small~\cite{jackson:1988,Sear1999}. The energetic contribution to the free energy is the energy of the fully-bonded system $-\beta\epsilon_{\mathrm{tot}}$, while the entropic term is approximated using the range of interaction and the width of the patches~\cite{Sear1999},
\begin{equation}
\beta\mu_s=\beta a_s=-3\ln(\lambda-1)-\ln\left(\frac{\delta^3}{\pi^2}\right)-\beta\epsilon_{\mathrm{tot}}.
\end{equation}

At phase coexistence, the temperature, pressure, and chemical potential of the fluid and solid phases have to be identical. The pressure of the solid is once again ignored, so the only remaining constraint is $\beta_{\mathrm{coex}}\mu_f=\beta_{\mathrm{coex}}\mu_s$. Using the equations above, it follows that
\begin{eqnarray}
\beta_{\mathrm{coex}} a_{\mathrm{HS}}+\beta a_{\mathrm{bond}}+\frac{\beta_{\mathrm{coex}} p_{\mathrm{HS}}}{\rho}+\frac{\beta_{\mathrm{coex}} p_{\mathrm{bond}}}{\rho}=\nonumber\\
=-3\ln(\lambda-1)-\ln\left(\frac{\delta^3}{\pi^2}\right)-\beta_{\mathrm{coex}}\epsilon_{\mathrm{tot}}.
\end{eqnarray}
As the hard-sphere system itself is temperature independent, it holds that
\begin{equation}
\beta_{\mathrm{coex}}\left(a_{\mathrm{bond}}+\frac{p_{\mathrm{bond}}}{\rho}+\epsilon_{\mathrm{tot}}\right)=C(\rho),
\end{equation}
where
\begin{equation}
C(\rho)=-\beta a_{\mathrm{HS}}-\frac{\beta p_{HS}}{\rho}-3\ln(\lambda-1)-\ln\left(\frac{\delta^3}{\pi^2}\right)
\end{equation}
is a function that only depends on $\rho$. It thus follows that $a_{\mathrm{bond}}+\frac{p_{\mathrm{bond}}}{\rho}+\epsilon_{\mathrm{tot}}$ represents a good temperature rescaling factor to obtain the master solubility line across the different models.

\begin{figure}[tb]
\begin{center}
\includegraphics[width=0.5\textwidth]{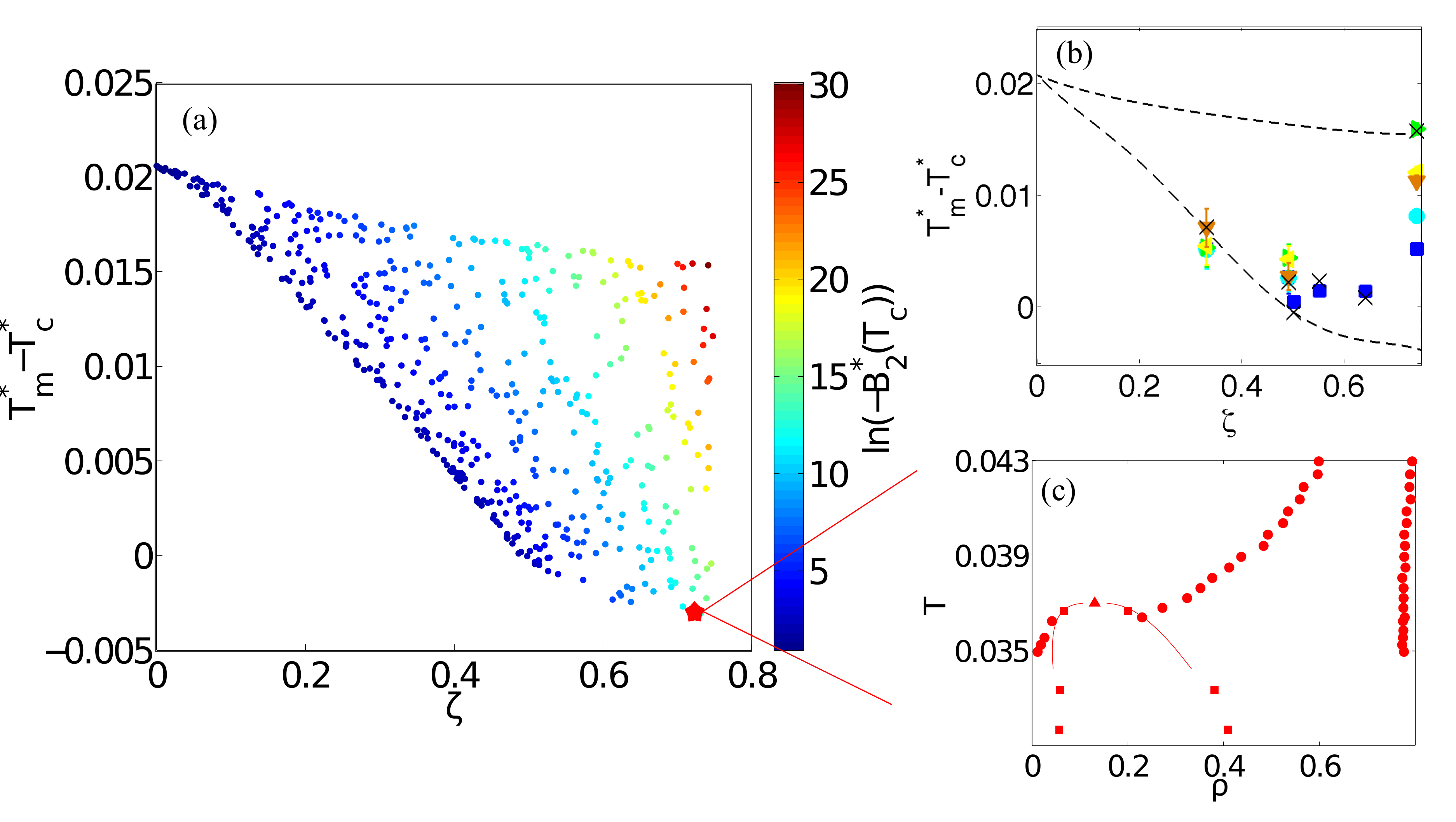}
\caption{(Color online) (a) WPT difference between rescaled melting temperature $T_m^*$ and rescaled critical temperature $T_c^*$ at the critical density $\rho_c\approx0.2$, i.e., the crystallization gap. Each circle represents a distinct $\{\epsilon_i\}$ realization and is colored by its  $\ln[-B_2(T_c)]$ value (higher values in the top right corner). (b) WPT predictions (black crosses) compared with simulation results for different patch geometries (symbols as in Fig.~\ref{fig:PD}(a)). The area within the dashed lines indicates the values covered by WPT in (a). (c) Phase diagram for $\{\epsilon_1=4.6655, \epsilon_2=1.2908, \epsilon_3=0.0437\}$ for which WPT predicts a stable gas-liquid coexistence. Even for an interaction range and patch coverage that would normally result into a metastable gas-liquid line, the bond energy asymmetry can lift $T_c$ above the solubility line (Appendix B).}\label{fig:slot}
\end{center}
\end{figure}

When compared with the simulated results, WPT overestimates the solubility temperature at all densities $\rho$, but nonetheless remarkably collapses the simulation results (Fig.~\ref{fig:PD}~(c) and (d)).  
The numerical validation of  WPT's $T_c$ predictions -- accurate to within 10-15\% -- allows us to estimate the size of the ``crystallization gap'' for a broader variety of models (Fig.~\ref{fig:slot})~\cite{footnote:4}. Interestingly, we find that for patch energy sets $\{\epsilon_i\}$ giving a same $\zeta$, a lower second virial coefficient $B_2$ results in a larger crystallization gap (Fig.~\ref{fig:slot}). Contrary to George and Wilson's (GW) crystallization slot proposal, i.e., that $\log(-\mathrm{B_2}^*)<5$ identifies facile crystallization~\cite{george:1994}, the asymmetric models reveal that $B_2$ does not by itself sets the size of the crystallization gap. The proposal is thus reasonable at low $\zeta$, but breaks down at high $\zeta$, where it even includes systems for which the critical point is fully stable (red star in Fig.~\ref{fig:slot} (a) and (c), Appendix B). In such systems, access to the crystal from a slowly concentrating, low-density solution would have to side-step the metastable gas-liquid coexistence regime. This regime typically prevents the formation of all but the smallest crystallites~\cite{vekilov:2002,Asherie2004}. High interaction asymmetry therefore provides a microscopic rationale for the failure of the GW proposal~\cite{bonnete:2001,fraden:2007}, which complements and supports previous suggestions that were based on a balance of specific/non-specific interactions~\cite{Haxton2012,Haxton2013}.

\section{Crystallization} 

Even if crystallization is thermodynamically possible the free energy drive may be insufficient to induce a phase transition on experimentally-relevant timescales. The role of asymmetry on homogeneous nucleation using umbrella sampling simulations is thus examined. 

We consider systems near their critical density $\rho_c\approx0.2$ at different degrees of supersaturation $\eta=\frac{T_m-T}{T_m-T_c}$,
where $T_m$ is the solubility temperature at that density. 
We determine the size of the crystal clusters in the simulation box following a standard procedure that defines \textit{crystal-like bond} and \textit{crystal-like particles}~\cite{saika:2011}. Due to the highly specific patch-patch interactions of our model, we generally define a crystal-like bond between particles $1$ and $2$ when they are actually bonded: $r<\lambda\sigma$, $\theta_{1,2i}<\delta$ and $\theta_{2,2i-1}<\delta$ for some $i$. A particle is considered to be crystal-like if it has six crystal-like bonds, and two crystal-like particles belong to the same crystal cluster if a crystal-like bond connects them.
Visual inspection of these ``crystals'' confirms that the criterion selects actual crystal clusters. 
For the umbrella sampling simulations, we use a biasing harmonic potential with spring constant $\kappa$ that varies between 0.07 and 0.12, depending on the model and the temperature studied
\begin{equation}
H^{\mathrm{bias}}=\kappa(s-s_0)^2,
\end{equation}
where $s$ is the size of the largest crystal cluster and $s_0$ is the target cluster size in the sampling window.
 Sampling windows are typically positioned every 3 particles, but denser sampling is sometimes required. The results of each simulation are then analyzed following a standard umbrella sampling protocol~\cite{saika:2011}.

\begin{figure}[tb]
\begin{center}
\includegraphics[width=0.5\textwidth]{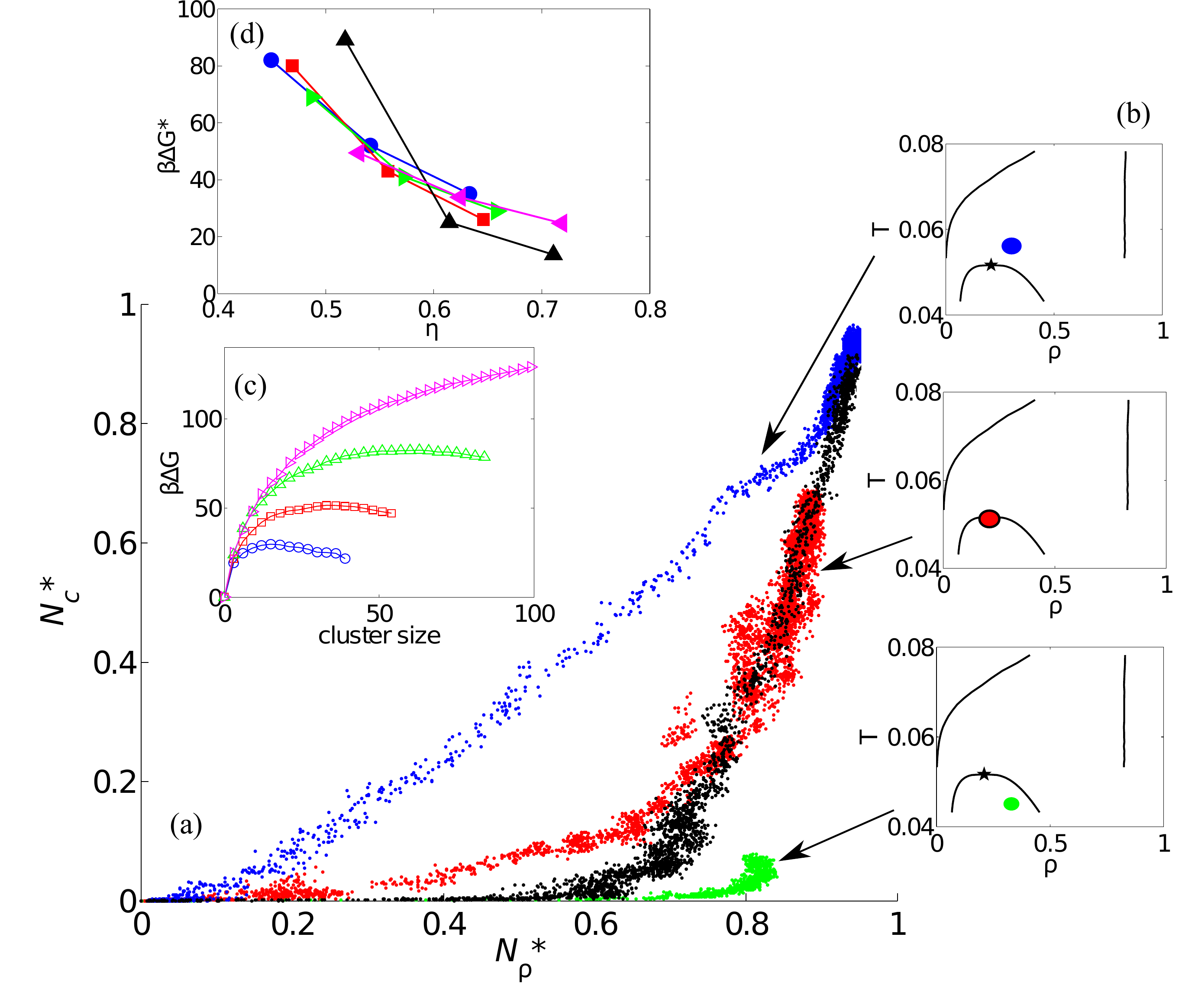}
\caption{(Color online) (a) Rescaled size of the largest liquid cluster $N_{\rho}^*$ compared to that of the largest crystal cluster $N_c^*$ for a model with $\zeta=0$ and $\gamma=0.017$ (blue/upper, red/dark gray and green/light gray) and $\zeta=0.2$ and $\gamma=0.017$ (black). The liquid cluster size is rescaled by the number of particles in the system ($N$=864) and the crystal cluster size by the size of the largest crystal cluster that fits in the simulation box (600). The different trends correspond to initial homogeneous fluid configurations under different conditions, as illustrated in (b). (c) Above the critical point, nucleation barriers can be computed  (blue circle $\eta=0.37$, red squares $\eta=0.46$, green triangles $\eta=0.55$ and magenta right-pointed triangles $\eta=0.64$). (d) Height of the nucleation barrier for different models (blue circles $\zeta=0$ and $\gamma=0.017$, red squares $\zeta=0$ and $\gamma=0.021$, green right-pointed triangles $\zeta=0$ and $\gamma=0.065$, black triangles $\zeta=0.1$ and $\gamma=0.017$, magenta left-pointed triangles $\zeta=0.2$ and $\gamma=0.017$.)}\label{fig:nucl_density}
\end{center}
\end{figure}

Unsurprisingly, the lower supersaturations correspond to higher free-energy barriers and larger critical nuclei (Fig.~\ref{fig:nucl_density} (c)). Across various patch geometries qualitatively similar results are obtained, but increasing the energy asymmetry significantly lowers the chemical potential difference, $\beta\Delta\mu$, between the fluid and the crystal. At high bond energy asymmetry fewer patches dominate the energy of the two phases, which makes that contribution in the two phases more similar and reduces the drive to crystallize. Higher densities are then needed to obtain a comparable nucleation barrier. Although this effect is not a fundamental limitation for particles to crystallize, real proteins in high-density solutions may partially unfold and aggregate, which interferes with their crystallization~\cite{vekilov:2002}. 
In addition, at high $\zeta$ the narrow crystallization gap results in larger free-energy nucleation barriers. High energy asymmetry thus hinders nucleation kinetics.

Classical nucleation theory (CNT) describes crystal formation reasonably well far above the critical point, but near and below $T_c$ the assembly behavior is more complex. Previous studies of isotropically attractive systems have shown that well below $T_c$ spinodal decomposition leads to dynamical arrest~\cite{Lu2008,Fortini2008}, because spontaneous density fluctuations result in dense regions within which binding is irreversible. In similar systems near the critical point, ``two-step'' nucleation is favored~\cite{Wolde1997}. Crystal formation is then more facile in high-density than in low-density fluid regions. The corresponding assembly behavior of patchy systems, whose low-density crystals may not be favored by spontaneous fluid density fluctuations~\cite{Haxton2012}, is here studied in unbiased constant $NpT$ MC simulations. These simulations sketch out the minimum free energy path for the assembly, which we track along the largest drop and the largest crystal cluster reaction coordinates (Fig.~\ref{fig:nucl_density} (a)).
The largest crystal cluster is determined as described above. Similarly, \textit{liquid-like particles} are defined as those that have at least four close neighbors (particles whose centers are within $\lambda\sigma$ of each other). Two \textit{liquid-like particles} belong to the same \textit{liquid cluster} if they are close neighbors. 
 These trajectories follow a fictitious dynamics without accounting for collective moves and where time should be properly rescaled. Yet, as previously showed~\cite{DeMichele2006}, such trajectories are representative of Brownian dynamics configurational space sampling for sufficiently short steps. The robustness of our observations is also confirmed by repeating the simulations using the virtual-move MC of Ref.~\cite{Whitelam2007}, which allows for collective rearrangements (details in Appendix C)~\cite{Whitelam2010,Haxton2012,Haxton2013}. We obtain for the symmetric case, $\zeta=0$, far above $T_c$, that the largest cluster formation is always crystalline and CNT applies. Near the critical point (within $\sim$10\% of $T_c$) a growing liquid drop first forms and only subsequent structural reorganization of the many micro crystals results in a large crystal cluster.

\begin{figure}[tbh]
\begin{center}
\includegraphics[width=0.45\textwidth]{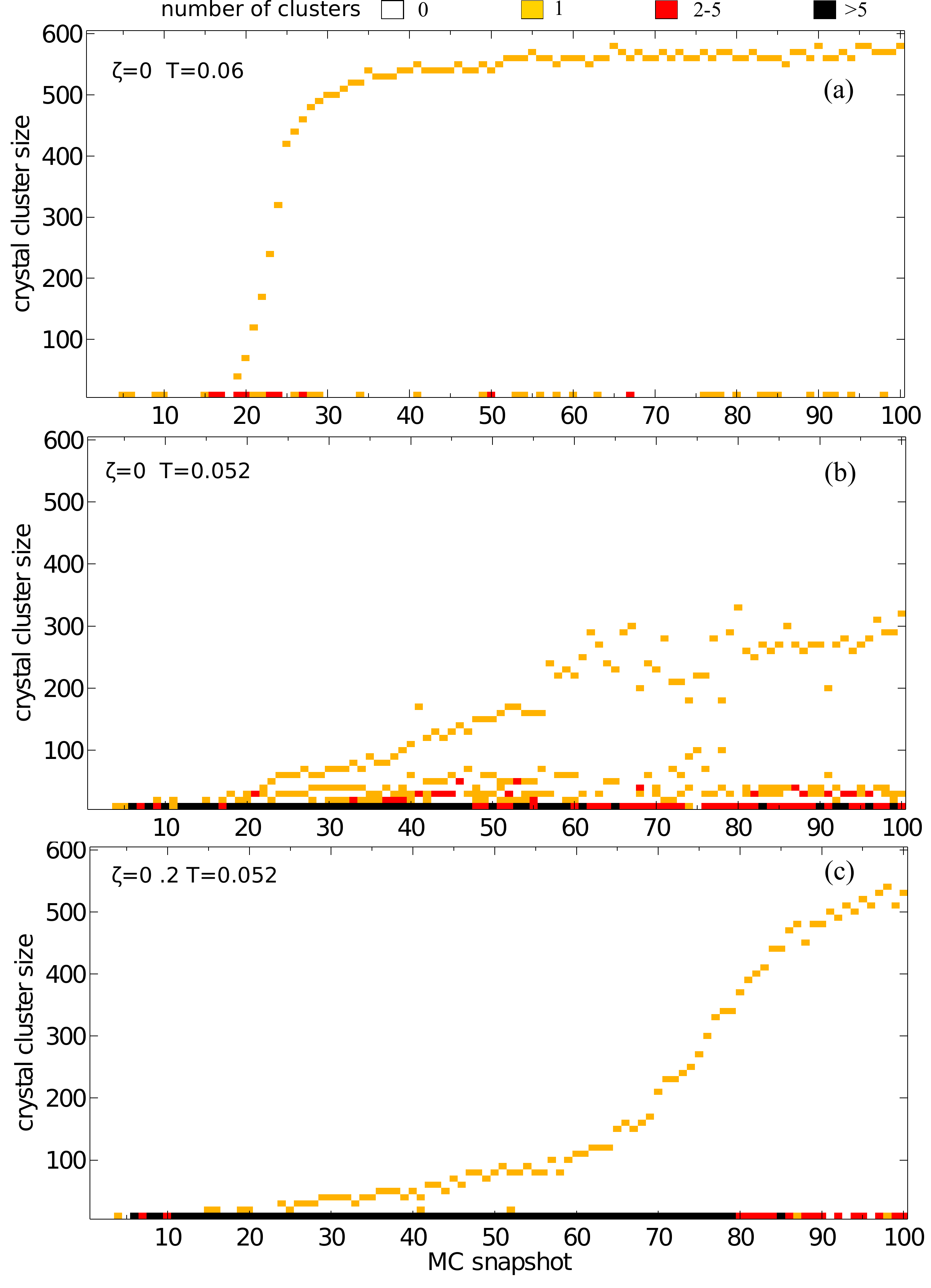}
\caption{(Color online) Distribution of crystal cluster sizes along the unbiased $NpT$ MC simulations respectively for (a) $\zeta=0$ at $T>T_c$, (b) $\zeta=0$ at $T=T_c$, and (c) $\zeta=0.2$ at $T=T_c$ for a system of 864 particles. The distribution of crystal cluster size from instantaneous snapshots is represented. White indicates the lack of cluster of that size, orange (lighter gray) one cluster being present, red (darker gray) between 2 and 5 clusters (few), and black indicates that more than 5 clusters are present (many). Five snapshots cover $10^5$ MC sweeps in the first two panels, and $5\times10^5$ in the last panel. }\label{fig:cl_distribution}
\end{center}
\end{figure}

 We can gain additional physical insights into the dynamical pathway by characterizing the distribution of crystal clusters within the largest liquid drop. Figure~\ref{fig:cl_distribution} reports the distribution of crystal cluster sizes from simulation snapshots. Panel (a) shows a classical nucleation scenario where, after a waiting time ($2\times10^6$ MC sweeps), a critical nucleus appears and grows rapidly. No secondary nucleation event is observed.
 Panel (b) illustrates the status of the system with symmetric interactions at the critical point. Almost instantaneously micro crystals (with fewer than 50 particles per cluster) form, and many of them survive the whole simulation. The formation of the largest cluster is much less smooth than in the classical nucleation limit. It is possible to see how the largest cluster breaks into two or more smaller aggregates and then forms again. The largest crystal cluster stems from the annealing of defects when multiple crystals assemble rather than from a single nucleation event.
 
Figure~\ref{fig:nucl_density} shows that the behavior at the critical point between low and higher energy asymmetry models is similar (black and red dots). Yet a closer analysis reveals that the cluster distribution exhibits a significant difference (Fig.~\ref{fig:cl_distribution}~(c)). In the asymmetric case, a single nucleation event is followed by the growth of a single crystal cluster rather than the re-organization of many micro crystals. Despite this resemblance with classical nucleation, the time between the appearance of a first critical nucleus and its full growth is long compared to a classical nucleation scenario in which nucleation is rare yet rapid. Crystallization occurs on a timescale similar to percolation and it is possible that the interplay between the two phenomena underlies the observed slow growth.
It is interesting to note that we do not observe any crystal cluster of significant size above the critical temperature within the simulation time even though the crystallization barrier height is similar to that of the symmetric case (Fig.~\ref{fig:nucl_density}(d)). This feature is left for future enquiries
 

\begin{figure}[tb]
\begin{center}
\includegraphics[width=0.5\textwidth]{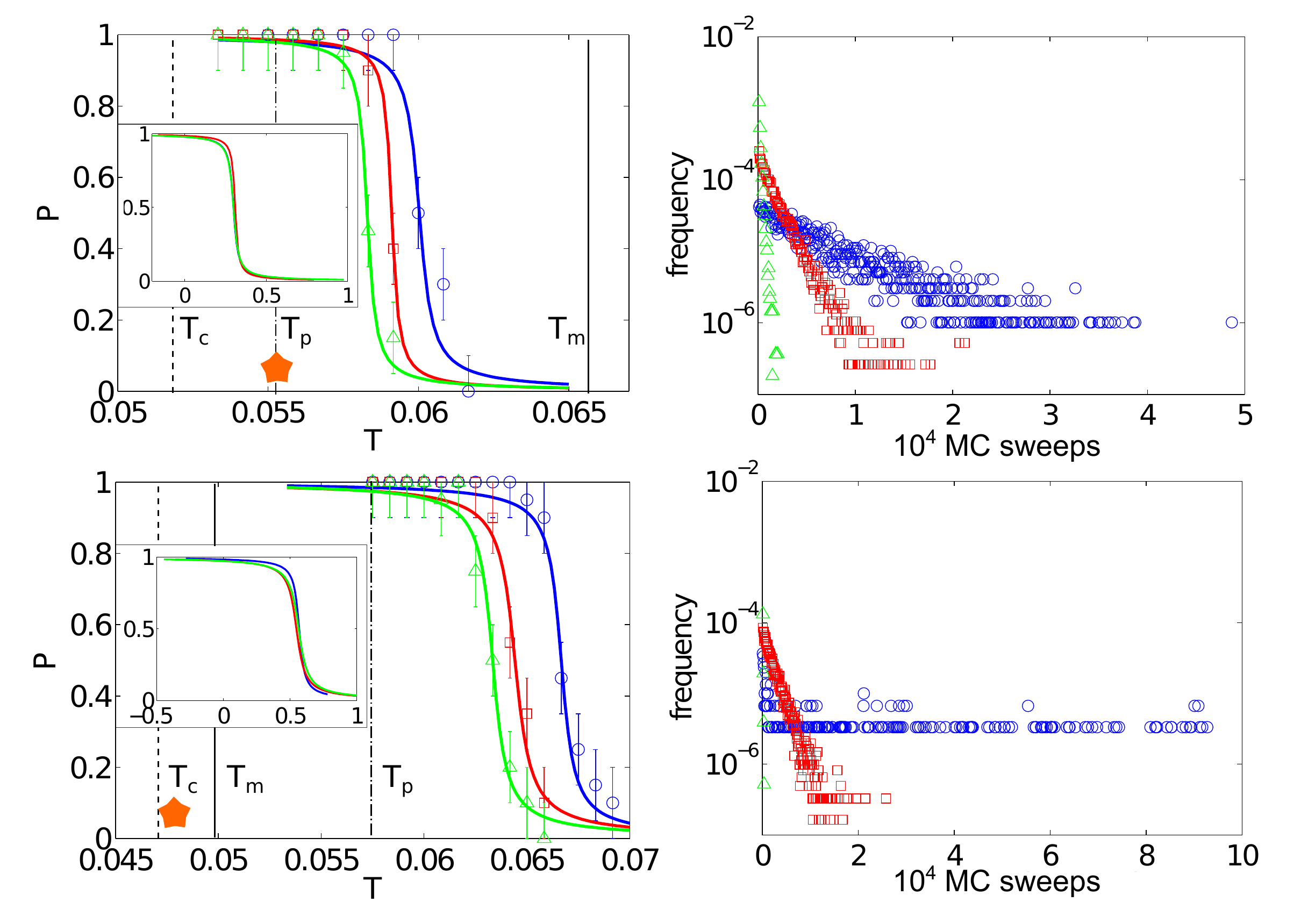}
\caption{(Color online) Percolation behavior for $\zeta=0.2$ (top) and $\zeta=0.55$ (bottom). Left panels show the probability 
$P$ of observing a spanning network as a function of temperature for system of size $N$=2048 (blue circles), 4000 (red squares) and 6912 (green triangles). The superimposed vertical lines indicate the critical temperature (dashed), the melting temperature (solid) and the percolation temperature (dot-dashed) estimated by finite-size scaling (inset). The right panels show the distribution of bond lifetime in the network at $T$=0.055 and $T$=0.048  respectively (orange stars in the left panels). Blue circles indicate the strongest bond, red squares the intermediate bond, and green triangles the weakest bond.}\label{fig:percolation}
\end{center}
\end{figure}

\section{Percolation} We finally consider whether direct percolation dynamically competes with crystallization. Below the percolation threshold $T_p(\rho)$ the system forms infinitely large spanning networks that can be long-lasting when bonding is strong~\cite{Zaccarelli2007,tavares:2010}. 

To explore the interplay between percolation and bond energy asymmetry (patch geometry asymmetry has but a weak impact), we determine $T_p(\rho_c)$ using finite-size rescaling~\cite{stauffer:1992}. 
We run 20 $NVT$ simulations with respectively $N=2048$, $4000$ and $6912$ for several temperatures at $\rho=0.2$. During the simulation, we determine the size of the biggest network defined as the largest set of particles connected by at least two bonds.  If such a cluster spans the whole simulation box along one dimension within $10^5$ MC sweeps, the system is deemed percolating. The percolation probability is the fraction of simulations showing such a percolating cluster. The data are in agreement with the tabulated 3D critical exponent to within 1\%~\cite{Lorenz1998,Klapp2013}.
For finite-size rescaling we use the tabulated critical exponents and the standard procedure~\cite{stauffer:1992}. 

Figure~\ref{fig:percolation} shows the results for systems with relatively low  ($\zeta=0.22$) and high ($\zeta=0.55$) asymmetry.
In the first system the percolation threshold lies just above $T_c$, while in the second, in which the strongest bond is much longer-lasting, $T_p$ is well above the solubility line. The dynamical relevance of percolation on crystallization is estimated from the distribution of bond lifetimes within the crystallization gap. At low bond energy asymmetry, the rearrangement of all bonds is observed within a few thousand MC steps. At high asymmetry, by contrast, the lifetime of the strongest bond (blue circles) is comparable to the length of the simulation ($10^5$ MC sweeps). The network is frozen, the bonds are almost irreversible, and no rearrangement takes place. This observation suggests that identifying the crystallization gap may be insufficient for crystallizing particles with high energy asymmetry, because a long-lasting gel caused by direct percolation then dynamically interferes with crystallization within the gap. Weakening the strongest bonds may be the only way to allow crystallization in these systems.

\section{Conclusion} In order to gain insights into protein crystallization and soft matter assembly more generally, we have considered the role of patch geometry and bond energy asymmetry on the crystal assembly of a family of schematic models. We find patch geometry asymmetry to have a weak effect, but bond energy asymmetry to severely impede the crystallization thermodynamics and kinetics. The crystallization gap shrinks, gel formation is favored, and nucleation shifts to higher supersaturations. The union of these observations suggests that to facilitate locating proper crystallization conditions, it is sometimes more effective to symmetrize the directional pair interactions between colloids or proteins, rather than specifically strengthen one of them, as is sometimes implicitly suggested~\cite{derewenda:2004,Fusco2012}. It also offers a rationalization of the GW crystallization slot proposal as well as of its occasional failure. At low bond energy asymmetry, the $B_2$ slot prescription falls within the slot; at high asymmetry, a large crystallization gap is only observed for $B_2$ below the slot, which corresponds to long-living gels, while for $B_2$ within the slot the crystallization gap is very small or even negative. The GW crystallization slot is therefore a necessary but insufficient condition for detecting optimal experimental conditions. 
 
Although we are now markedly closer to understanding simple, monomeric protein crystallization, the assembly features of more complex proteins remain a challenge. Some proteins dimerize or change conformation on a timescale comparable to their crystallization, while membrane proteins typically require entirely different crystallization approaches than the one considered in this work. Further modifications to patchy particle models, such as self-interacting or dynamically evolving patches, may thus guide our understanding of these complex yet crucial molecules.

\begin{acknowledgments}
PC acknowledges NSF support No.~NSF DMR-1055586.
\end{acknowledgments}

\appendix

\section{Saddle point approximation}
The analytical expression for $f(\boldsymbol{\alpha})$ in Eq.~\ref{eq:saddle_point} is
\begin{widetext}
\begin{align}
f(\boldsymbol \alpha)&=\frac{1}{2}\left\{4-3\sqrt{1-y^2}\cos(\phi)\cos(\phi_0)\sin(\theta_0)-\sqrt{1-y^2}\cos(\chi)\cos(\chi_0)\sin(\theta_0)+\sqrt{1-y^2}\cos(\chi)\cos(\chi_0)\cos^2(\zeta)\sin(\theta_0)\right.\nonumber\\
&-\sqrt{1-y^2}\cos(\phi)\cos(\phi_0)\cos(2\zeta)\sin(\theta_0)-2\sqrt{1-y^2}\cos(\phi)\cos(\zeta)\sin(\phi_0)\sin(\chi_0)\sin(\zeta)\nonumber\\
&-2y\cos(\chi_0)\cos(\zeta)\sin(\theta_0)\sin(\zeta)+2y\cos(\phi)\cos(\chi)\sin(\phi_0)\sin(\chi_0)\sin^2(\zeta)-2\cos(\phi)\cos(\phi_0)\sin(\chi)\sin(\chi_0)\sin^2(\zeta)\nonumber\\
&-\sqrt{1-y^2}\cos(\chi)\cos(\chi_0)\sin(\theta_0)\sin^2(\zeta)+y\cos(\phi)\cos(\phi_0)\cos(\chi)\sin(\theta_0)\sin(2\zeta)+\cos(\phi)\sin(\phi_0)\sin(\chi)\sin(\theta_0)\sin(2\zeta)\nonumber\\
&+\cos(\theta_0)\left[-3y+\cos(\phi_0)\cos(\chi_0)\sin(\phi)\sin(\chi)+\cos^2(\zeta)\left(-y+y\cos(\chi)\cos(\chi_0)\sin(\phi)\sin(\phi_0)\right.\right.\nonumber\\
&\left.-\cos(\phi_0)\cos(\chi_0)\sin(\phi)\sin(\chi)\right)-2\sqrt{1-y^2}\cos(\chi)\cos(\zeta)\sin(\zeta)+y\sin^2(\zeta)+\cos(\phi_0)\cos(\chi_0)\sin(\phi)\sin(\chi)\sin^2(\zeta)\nonumber\\
&-2\cos(\phi)\cos(\chi_0)\sin(\phi_0)\sin(\chi)\sin^2(\zeta)-y\cos(\chi)\cos(\chi_0)\left(2\cos(\phi)\cos(\phi_0)\sin^2(\zeta)+\sin(\phi)\sin(\phi_0)(1+\sin^2(\zeta))\right)\nonumber\\
&\left.+\sqrt{1-y^2}\cos(\phi)\cos(\phi_0)\cos(\chi_0)\sin(2\zeta)+\sqrt{1-y^2}\cos(\chi_0)\sin(\phi)\sin(\phi_0)\sin(2\zeta)\right]\nonumber\\
&+\sin(\phi)\left[-\sin(\phi_0)\left(2\sin(\chi)\sin(\chi_0)\sin^2(\zeta)+\sin(\theta_0)\left(3\sqrt{1-y^2}+\sqrt{1-y^2}\cos(2\zeta)-2y\cos(\chi)\cos(\zeta)\sin(\zeta)\right)\right)\right.\nonumber\\
&\left.\left.+\cos(\phi_0)\left(-2\cos(\zeta)\sin(\chi)\sin(\theta_0)\sin(\zeta)+\sin(\chi_0)\left(-2y\cos(\chi)\sin^2(\zeta)+\sqrt{1-y^2}\sin(2\zeta)\right)\right)\right]\right\}.\nonumber
\end{align}
\end{widetext}

\section{Grand Canonical MC approximation for critical point}
\begin{figure}[bth]
\begin{center}
\includegraphics[width=0.4\textwidth]{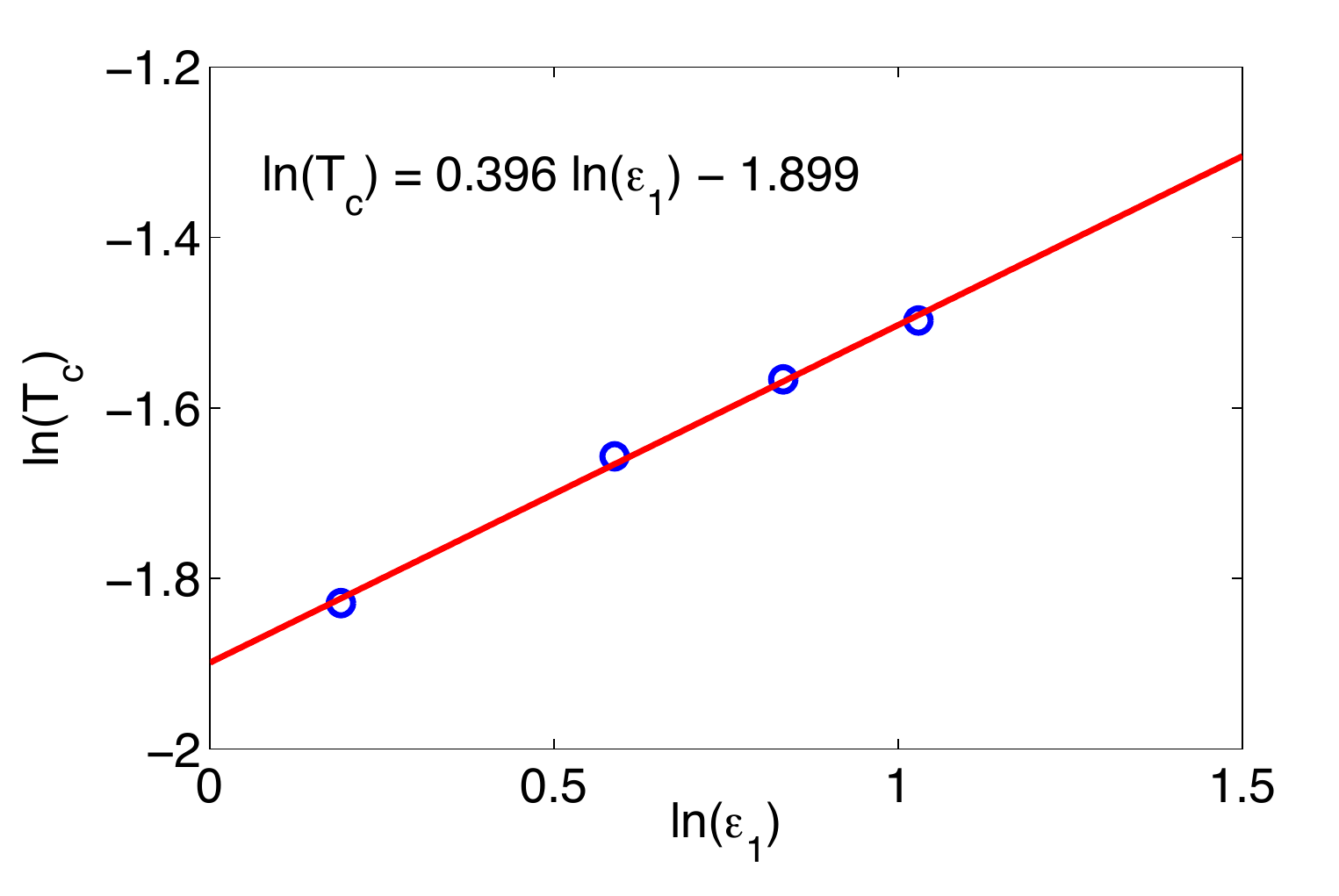}
\caption{(Color online) Fit of the critical point determined with GCMC simulations for systems with an increasingly strong strongest patch.}\label{fig:GCMC}
\end{center}
\end{figure}

To check the position of the critical point for the extreme case of Fig.~\ref{fig:slot} (c) ($\epsilon_1=4.6655$, $\epsilon_2=1.2908$ and $\epsilon_3=0.0437$), we perform grand canonical MC (GCMC) simulations. Because one of the patches is markedly stronger than the others, percolation takes place at a relatively high temperature. Consequently, GCMC samples slowly and poorly, and the critical temperature estimate is affected by large errors. To obtain a better estimate of the phase diagram, we perform GCMC for systems with an increasing strength of the strongest patch (keeping the other patches identical) and we fit a power-law to the resulting critical temperatures (Fig.~\ref{fig:GCMC}). The value of the fit for $\epsilon_1=4.6655$ ($T=0.045$ in units of $\epsilon_{\mathrm{tot}}$) confirms the stability of the critical point with respect to the solubility line.

\section{Virtual move MC}

\begin{figure}[tbh]
\begin{center}
\includegraphics[width=0.5\textwidth]{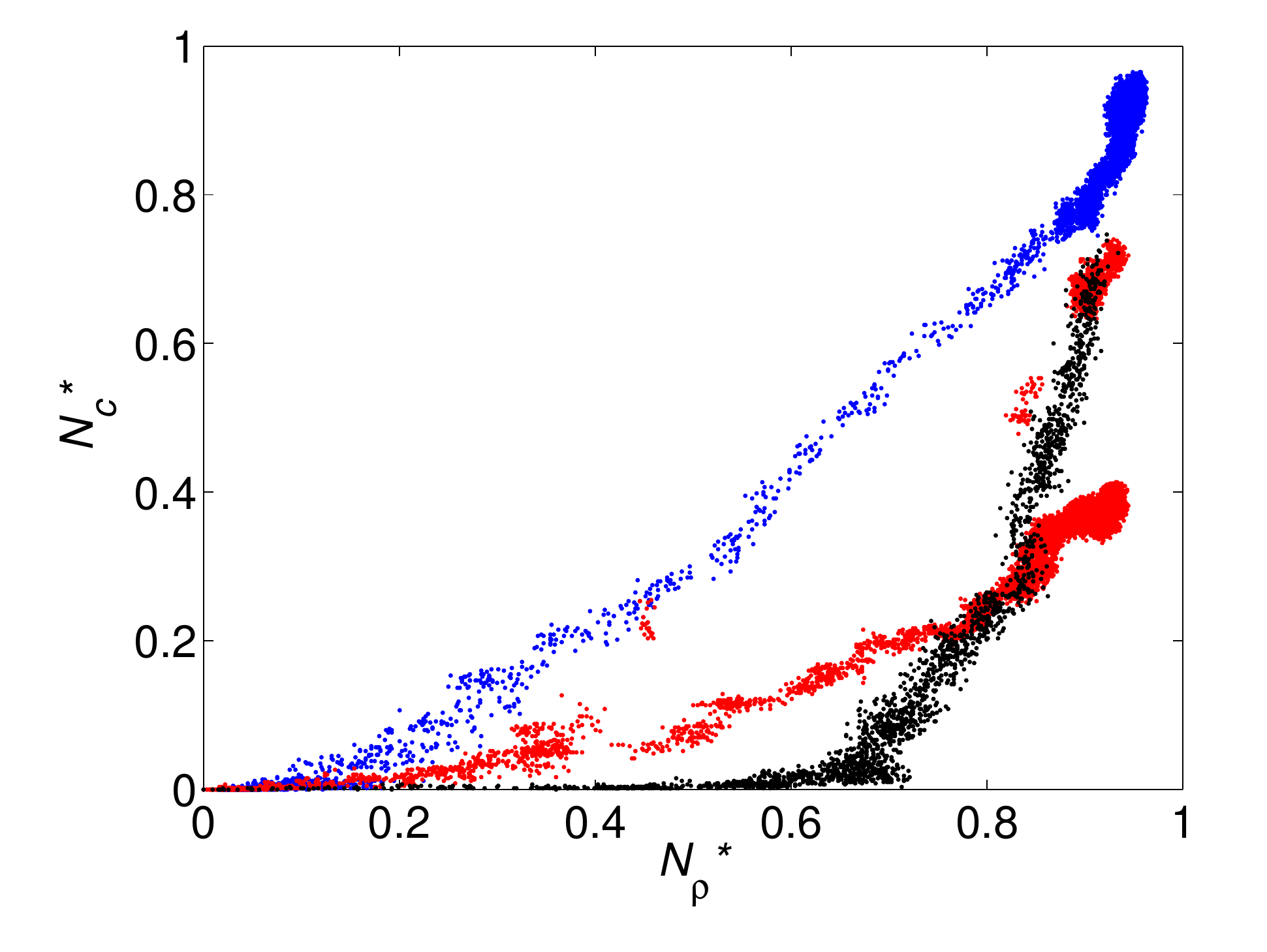}
\caption{(Color online) Crystallization pathways as in Fig.~3(a) using virtual move MC to simulate the dynamics. Comparison with Fig.~3(a) indicates that the crystallization pathways do not depend on the microscopic dynamics.}\label{fig:Whitelam}
\end{center}
\end{figure}

Standard MC simulations are based on sequential perturbation of the system and do not directly account for the collective moves through which the system sometimes relaxes. Although it has been shown that for small enough displacements, MC recovers the Brownian dynamics of patchy particle models~\cite{DeMichele2006}, it is reasonable to wonder if collective moves could nonetheless affect the system's dynamics. To check this possibility, we implement the virtual move MC algorithm~\cite{Whitelam2007}, which accommodates cluster displacement and rotations and prevents the system to be stuck in unphysical traps. This commonly used algorithm has been shown to reproduce real dynamics of short-range attractive systems and it is commonly used for this purpose~\cite{Whitelam2007,Whitelam2010,Haxton2012,Haxton2013}.

A virtual move consists of identifying a cluster to randomly displace or rotate. Each displacement draws from a uniform distribution between 0 and 0.2$\sigma$ and each rotation uniformly selects an axis of rotation and an angle of rotation. Following Ref.~\cite{Whitelam2007}, to avoid generating large clusters whose moves will often be rejected, we draw the cutoff $n_c$ of the cluster size from $P(n_c)\propto n_c^{-1}$. 
The results in Fig.~\ref{fig:Whitelam} are in agreement with those generated by the standard $NpT$ MC (Fig.~\ref{fig:nucl_density}), and confirm the robustness of the phenomenology with respect to changes in the microscopic dynamics.

\bibliographystyle{prsty}
\bibliography{MD_paper}

\begin{thebibliography}{10}
\bibitem{footnote:1}
The unit temperature $T$ (and its inverse, $\beta$) is set by
  $\epsilon_1+\epsilon_2+\epsilon_3=\epsilon_{\mathrm{tot}}$ with Boltzmann's
  constant $k_B=1$. The geometrical orientation varies from one pair of patches
  to the next. Although these models can crystallize in other lattices at high
  temperatures and densities, these structures are not relevant for protein
  crystallization.
\end{thebibliography}

\end{document}